\documentclass[12pt]{iopart}

\usepackage{iopams}  
\usepackage{graphicx}
\begin{document}

\title[Bounding network spectra]{Bounding network spectra for network design}

\author{Adilson E. Motter}

\address{Department of Physics and Astronomy and Northwestern Institute on
Complex Systems (NICO), Northwestern University, Evanston, IL 60208, USA}
\ead{motter@northwestern.edu}

\begin{abstract}

The identification of the limiting factors in the dynamical behavior of 
complex systems is an important interdisciplinary problem which often can 
be traced to the spectral properties of an underlying network. By deriving 
a general relation between the eigenvalues of weighted and unweighted
networks, here I show that for a wide class of networks the dynamical
behavior is tightly bounded by few network parameters. This result
provides rigorous conditions for the design of networks with predefined
dynamical properties and for the structural control of physical processes 
in complex systems.  The results are illustrated using synchronization
phenomena as a model process. 

\end{abstract}

\vspace{2pc}

\submitto{\NJP}
\maketitle


\section{Introduction}

Complex dynamical systems are high dimensional in nature. The determination
of simple general principles governing the behavior of such systems is an
outstanding problem which has attracted a great deal of attention in 
connection with recent network and graph-theoretical constructs \cite{physicad,nbw:2006}.
Here I focus on synchronization, which is the process that has attracted most
attention, and use this process to study the interplay between network
structure and dynamics. 
Synchronization is a widespread phenomenon in distributed systems, with examples
ranging from neuronal to technological networks \cite{S:2003}. 
Previous studies have shown that network synchronization is strongly influenced
by the randomness \cite{W:1999,BP:2002}, degree (connectivity) distribution 
\cite{NMLH:2003}, correlations \cite{MCK3,BGS:2005}, and distributions of
directions and weights \cite{MCK1,MCK2} in the underlying network of couplings. 
But what is the ultimate origin of these dependences?

In this paper, I show that these and other important effects in the dynamics of
complex networks are ultimately controlled by a small number of network parameters. 
For concreteness, I focus on complete synchronization of identical dynamical units \cite{msf},
which has served as a prime paradigm for the study of collective dynamics in complex networks.
In this case, the synchronizability of the network is determined by the
largest and smallest nonzero eigenvalues of the coupling (Laplacian) matrix. My principal
result is that, for a wide class of complex networks, these eigenvalues are
tightly bounded by simple functions of the weights and degrees in the network.
The quantities involved in the bounds are either known by construction or can
be calculated in at most $O(kN)$ operations for networks with $N$ nodes and
$2kN$ links, whereas the numerical calculation of the eigenvalues of large
networks would be prohibitively costly since it requires in general 
$O(N^3)$ operations even for the special case of undirected networks. These 
bounds are in many aspects different from those known in the literature of
graph spectral theory \cite{mohar} and are suitable to relate the physically
observable structures in the network of couplings to the dynamics of the
entire system.

The eigenvalue bounds are then applied to design complex networks that display
predetermined dynamical properties and, conversely, to determine how given 
structural properties influence the network dynamics. This is achieved
by exploring the fact that the quantities used to express the bounds have
direct physical interpretation. This leads to conditions for the enhancement
and suppression of synchronization in terms of physical
parameters of the network. The main results also apply to a class of weighted
and directed networks and are thus important to
assess the effect of nonuniform connection weights in the synchronization of
real-world networks \cite{BBPV:2004}. The proposed method for network
design is based on a relationship between the eigenvalues of a substrate network
that incorporates the structural constraints imposed to the system and those of
weighted versions of the same network. This method is thus complementary to
other recently proposed approaches for identifying  \cite{TM2,JT06,YRK06}
or constructing \cite{TM1,MT06,MT062} networks with desired dynamical properties. 

The paper is organized as follows. In Sec.\ \ref{sec2}, I define the class of
networks to be considered and announce the main result on the eigenvalue
bounds, which is proved in the appendix. In Sec.\ \ref{sec3}, I discuss 
an eigenvalue approach to the study of network synchronization. The problem of 
network design and the impact of the network structure on dynamics is considered 
in Secs.\ \ref{sec4} and \ref{sec5}, respectively. 
Concluding remarks are incorporated in the last section. 

\section{Eigenvalue Bounds}
\label{sec2}

The dynamical problems considered in this paper are related to the
extreme eigenvalues of the Laplacian matrix. This section concerns
the bounds of these eigenvalues.

\subsection{Class of Networks}

Most previous studies related to network spectra and dynamics
have focused on unweighted networks of symmetrically coupled nodes. 
In order to account for some important recent models of weighted
and directed networks, here I consider a more general class of networks. 
The networks are
defined by adjacency matrices $A$ satisfying the condition that
\begin{equation}
\hat{A}=\left(\frac{k_i}{S_i}A_{ij}\right), \;\;\; i,j =1,\dots , N,
\label{e3}
\end{equation} 
is a symmetric matrix,  where $k_i\ge 1$ is the degree of node $i$, factor
$S_i=\sum_j A_{ij}>0$ is the total strength of the input connections at node
$i$, and $N$ is the number of nodes in the network. 
According to this condition, the
in- and out-degrees
are 
equal at each node of the network, although the strengths
of in- and out-connections are not necessarily the same. Matrix
$\hat{A}=(\hat{A}_{ij})$ is possibly weighted: 
$\hat{A}_{ij}>0$ if there is a connection between nodes $i$ and $j\ne i$
and $\hat{A}_{ij}=0$ otherwise, where $\sum_j \hat{A}_{ij}=k_i$ because of the
normalization factor $S_i/k_i$. 
The class of networks defined by Eq.~(\ref{e3}) includes
as particular cases all undirected networks (both unweighted and weighted) and 
all directed networks derived from undirected networks by a node-dependent
rescaling of the input strengths. The dominant directions of the couplings are
determined by $S_i/k_i$ and the weights by both $S_i/k_i$ and $\hat{A}$,
where $S_i/k_i$ defines the mean and $\hat{A}$ the relative strength of the
individual input connections at node $i$. The usual unweighted undirected 
networks correspond to the case where $\hat{A}$ is binary and $S_i=k_i$ for
all the nodes. 

The study of this class of networks is motivated by both physical and mathematical 
considerations.
From the mathematical viewpoint, I show in the appendix that the conditions imposed to
matrix $A$ guarantee that the corresponding coupling matrices are diagonalizable and have
real spectra. Physically, this coupling scheme is general enough to reproduce the weight
distribution of numerous realistic networks \cite{BBPV:2004} and to show how the combination
of topology, weights, and directions affect the dynamics. 
Indeed, the weighted and directed networks comprised by the adjacency matrix $A$ in
(\ref{e3}) include important models previously considered in the literature,
such as the models where $S_i=1$ $\,\forall i$, used to study coupled maps
\cite{jj02,ja03} and to address the effects of asymmetry and saturation of
connection strengths \cite{MCK1,MCK2}. It also includes the models introduced
in Refs.~\cite{MCK3,lgh2004,zmk06}, where the connection strengths depend on
the degrees of the neighboring nodes, and other models reviewed in Ref.~\cite{b:2006}.
In what follows, I consider the general class of networks defined by
Eq.~(\ref{e3}) with the additional assumption that each network has a single
connected component.

\subsection{Coupling Matrices}

The coupling matrix relevant to this study is the Laplacian matrix
$G=(G_{ij})$, where
\begin{equation}
G_{ij}=\left(\delta_{ij}S_i- A_{ij}\right).
\label{e5}
\end{equation}
The Laplacian matrix can be written as $G=S\hat{G}=SD^{-1}L$, where 
$S=(\delta_{ij}S_i)$ is the matrix of input strengths,  
$\hat{G}=(L_{ij}/k_i)$ is a normalized Laplacian matrix,
$D=(\delta_{ij}k_i)$ is the matrix of degrees, and
$L=(\delta_{ij}k_i-\hat{A}_{ij})$.
As shown in the appendix, matrices $G$ and $\hat{G}$ are diagonalizable
and all the eigenvalues of $G$ and $\hat{G}$ are real.
For connected networks where all the input strengths $S_i$ are positive, 
as assumed here, the eigenvalues of matrices
$G$, $\hat{G}$, and $S$ can be ordered as
\begin{eqnarray}
0=\lambda_1 < \lambda_2 &\le& \cdots \le \lambda_N,
\label{eq31}\\
0=\mu_1 <\mu_2 &\le& \cdots \le \mu_N,
\label{eq32}\\
0<\nu_1 \le \nu_2 &\le& \cdots \le \nu_N,
\label{eq33}
\end{eqnarray} 
respectively. The strict inequalities $\lambda_2>0$ and $\mu_2>0$ follow from
Eq.~(\ref{eeq9}), which expresses $\lambda_2$ (and also $\mu_2$ if one takes $S_i=1$ $\forall i$)
as a sum of nonnegative terms with at least one of them being nonzero when the network is connected.
The identities $\lambda_1=\mu_1=0$ are a  simple consequence
of the zero row sum property of matrices $G$ and $\hat{G}$.

\subsection{Extreme Eigenvalues: Bounds for Arbitrary Network Structure}

I now turn to the analysis of the eigenvalues of the Laplacian matrix $G$. I use
$k_{\min}$, $k_{\max}$ and $k$ to denote the minimum, maximum and mean degree
in the network. The minimum and maximum input strengths are denoted by 
$S_{\min}=\min_{i}\{S_i\}$ and $S_{\max}=\max_{i}\{S_i\}$, respectively,
while $k_{\min^*}$ is used to denote the minimum degree among the nodes with input strength
$S_{\min}$. I first state the following general theorem.
\medskip

\noindent{\bf Theorem:} The largest and smallest nonzero eigenvalues of matrices
$G$, $\hat{G}$, and $S$ are related as
 \begin{eqnarray}
\nu_N &\le&\lambda_N\le \nu_N\mu_N,
\label{eq41}\\
\nu_1\mu_2 h &\le&\lambda_2 \le \nu_1 g,
\label{eq42}
\end{eqnarray}
where $h=(\sum_i k_i/S_i)/\sqrt{(\sum_i k_i)(\sum_i k_i/S_i^2)}$, 
$g=(1-\beta)^{-1}$, and 
$\beta=(k_{\min^*}/S_{\min})/(\sum_i k_i/S_i)$, for any network with adjacency
matrix satisfying (\ref{e3}).
\medskip

This theorem is important because it relates the desired 
and usually unknown eigenvalues of Laplacian matrix $G$ with the input strengths and
the often approximately known eigenvalues of the normalized Laplacian matrix $\hat{G}$.
In general, one has $\mu_2\le N/(N-1)\le\mu_N\le 2$, which follows as a
simple generalization of the results in Ref. \cite{chung:book} to the
weighted and directed networks defined by Eq.~(\ref{e3}). Physically, the eigenvalues
$\mu_2$ and $\mu_N$ are related to relaxation rates \cite{MCK2}, while $\nu_1$ and $\nu_N$
are just the input strengths $S_{\min}$ and $S_{\max}$, respectively.
A special case of the theorem was announced in Ref.~\cite{zmk06}.
The theorem is proved in the appendix. In the remaining part of the paper I explore
applications of the theorem.

\section{Synchronization Problem}
\label{sec3}

In this section, networks of identical oscillatory systems are used to
discuss how the  coupling cost and stability of synchronous states are
expressed in terms of the eigenvalues considered in the previous section.

\subsection{Oscillator Network}

Consider a network of $N$ diffusively coupled dynamical units \cite{msf} modeled by
\begin{equation}
\dot{\bf x}_i = {\bf F}({\bf x}_i)-\sigma\sum_{j=1}^{N} G_{ij} {\bf H}({\bf x}_j), \;\;\; i=1,\dots , N,
\label{e2}
\end{equation}
where the first term on the r.h.s.~describes the dynamics of each unit, while the second
equals $\sigma\sum_{j=1}^{N}A_{ij}\left[{\bf H}({\bf x}_j) - {\bf
H}({\bf x}_i) \right]$ and
accounts for the couplings between different units: ${\bf H}({\bf x}_j)$ is the signal
function that describes the influence of unit $j$ on the units coupled to $j$ and
$\sigma \ge 0$ is the overall coupling strength. The adjacency matrix
$A=(A_{ij})$ satisfies (\ref{e3}) and is related to the  Laplacian matrix
$G=(G_{ij})$ through Eq.~(\ref{e5}). 

Completely synchronized states $\{{\bf x}_i(t)={\bf s}(t),\;\forall i\; |\; \dot{\bf s}={\bf F}({\bf s})\}$
are always solutions of system (\ref{e2}). Since the Laplacian matrix is diagonalizable,
the stability of these synchronous states can be studied using the standard master
stability framework \cite{msf} (see also \cite{TM2,TM1}). This reduces the
variational equations of system (\ref{e2}) to $N$  blocks of the form
$\dot{\bf y}_i=[D{\bf F}({\bf s})-\sigma \lambda_i D{\bf H}({\bf s})]{\bf y}_i$,
where ${\bf y}_2,\dots ,{\bf y}_N$ correspond to perturbations transverse to 
the synchronization manifold. The synchronous state ${\bf s}$ is linearly
stable if and only if the largest Lyapunov exponent $\Lambda(\sigma \lambda_i)$
for this equation is negative for each transverse mode $i=2,\dots , N$, where
$\{\lambda_i\}_{i=2}^N$ are the nonzero eigenvalues of $G$ in Eq.\ (\ref{eq31}).

\subsection{Stability and Coupling Cost}

In a broad class of oscillatory dynamical systems, function $\Lambda$ is negative
in a single interval $(\alpha_1,\alpha_2)$ \cite{BP:2002,MCK2,msf}. The synchronous state
is then stable
for some $\sigma$ if the eigenvalues of the Laplacian matrix $G$ satisfy the
condition \cite{BP:2002}
\begin{equation}
R[G]\equiv \frac{\lambda_N}{\lambda_2} < \frac{\alpha_2}{\alpha_1}[{\bf F},{\bf H},\bf{s}].
\label{e7}
\end{equation}
The r.h.s.~of this inequality depends only on the dynamics while the l.h.s.~depends only
on the structure of the network, as indicated in the brackets. The smaller the ratio of
eigenvalues $R$ the larger the number of dynamical states for which condition (\ref{e7})
is satisfied. 
Moreover, when this condition is satisfied and $\alpha_2/\alpha_1$ is finite,
the smaller the ratio $R$ the larger the relative interval of the coupling
parameter $\sigma$ for which the corresponding synchronous state is stable.

When condition (\ref{e7}) is satisfied, the eigenvalues $\lambda_2$ and
$\lambda_N$ are related to the synchronization thresholds as
\begin{eqnarray}
\label{e71}
\lambda_2 &=& \alpha_1/\sigma_{\min},\\
\lambda_N &=& \alpha_2/\sigma_{\max},
\label{e72}
\end{eqnarray}
where $\sigma_{\min}$ and $\sigma_{\max}$ are the minimum and maximum coupling
strengths for stable synchronization, respectively. These relations will be explored in the
design of networks with predefined thresholds in Sec.~\ref{sec4}.

This characterization is not complete without taking into account the cost
involved in the coupling. The coupling cost required for stable synchronization
was defined in Refs.\ \cite{MCK1,MCK2} as the sum of the coupling strengths at 
the lower synchronization threshold,
$C\equiv\sigma_{\min}\sum_{i,j}A_{ij}= \alpha_1/\lambda_2 \sum_{i}S_{i}$. 
This cost function can be expressed in terms of eigenvalues of the Laplacian 
matrix \cite{TM1}, \begin{equation}
C = \frac{\alpha_1}{\lambda_2}\sum_{j=2}^{N}\lambda_j,
\label{e8}
\end{equation}
and depends separately on the dynamics ($\alpha_1$) and structure ($\sum_j \lambda_j/\lambda_2$)
of the network. This can be used to derive an upper bound for $C$ expressed in
terms of the ratio $R$:
\begin{equation}
\alpha_1 (N-1)\le C\le \alpha_1 (N-1)\, R.
\label{e9}
\end{equation}
Therefore, ratio $R$ is a measure of the synchronizability {\it and} cost of the
network, with the interpretation that the network is more synchronizable and the
cost is more tightly upper bounded when $R$ is smaller. 
The synchronization problem is then reduced to the study of eigenvalues of the Laplacian 
matrix $G$.

\section{Design of Networks with Predefined Synchronization Thresholds}
\label{sec4}

In this section, I show how the theorem of Sec.\ \ref{sec2} can be used to 
design large networks with predetermined eigenvalues $\lambda_2$ and $\lambda_N$.
In the synchronization problem of Sec.~\ref{sec3}, this corresponds to the
design of networks with predetermined lower 
($\sigma_{\min}=\alpha_1/\lambda_2$) and upper ($\sigma_{\max}=\alpha_2/\lambda_N$)
synchronization thresholds.

\subsection{Network Design}

\begin{figure}[t]
\begin{center}
\includegraphics[width=0.48\textwidth]{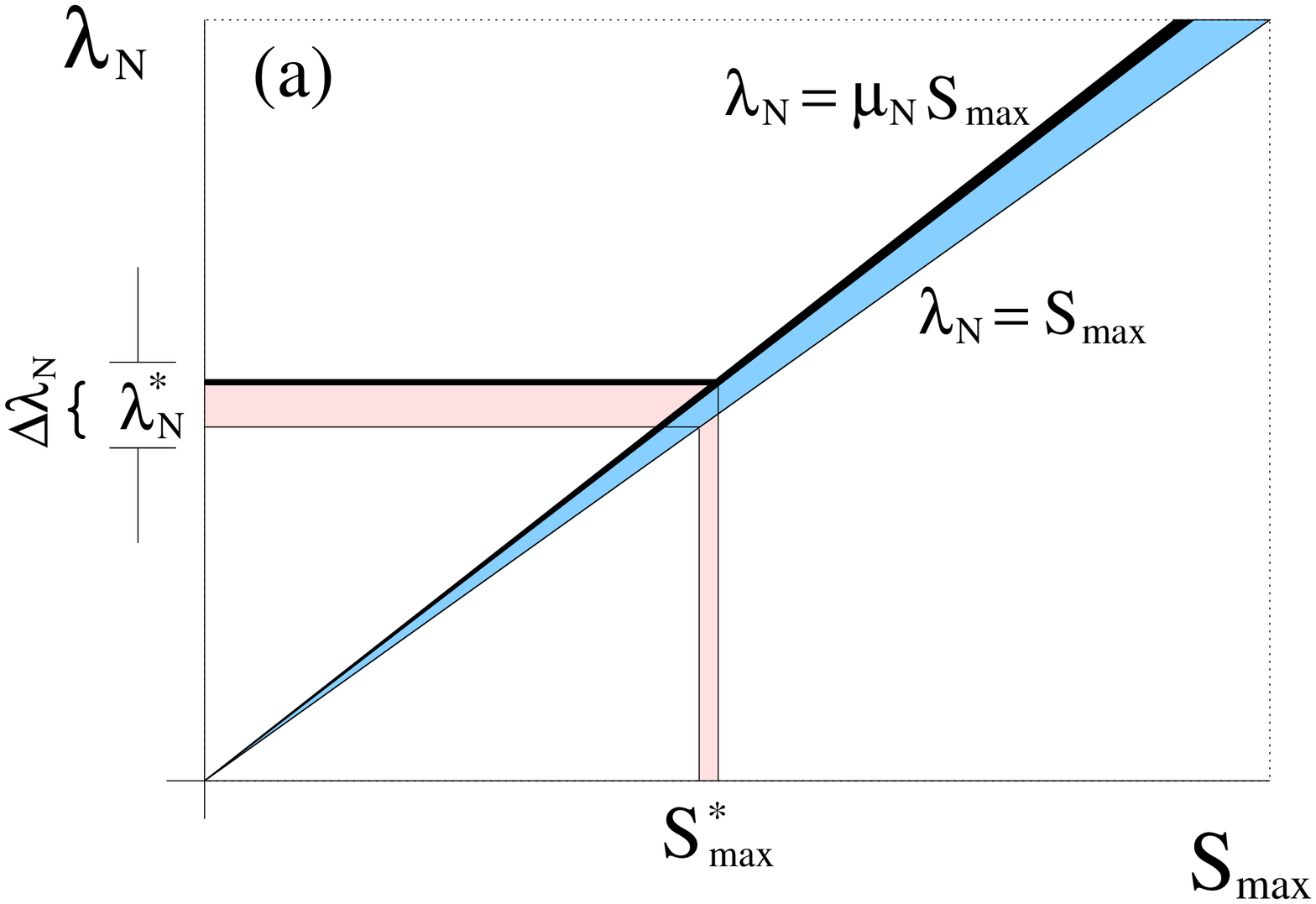}\hspace{0.5cm}
\includegraphics[width=0.48\textwidth]{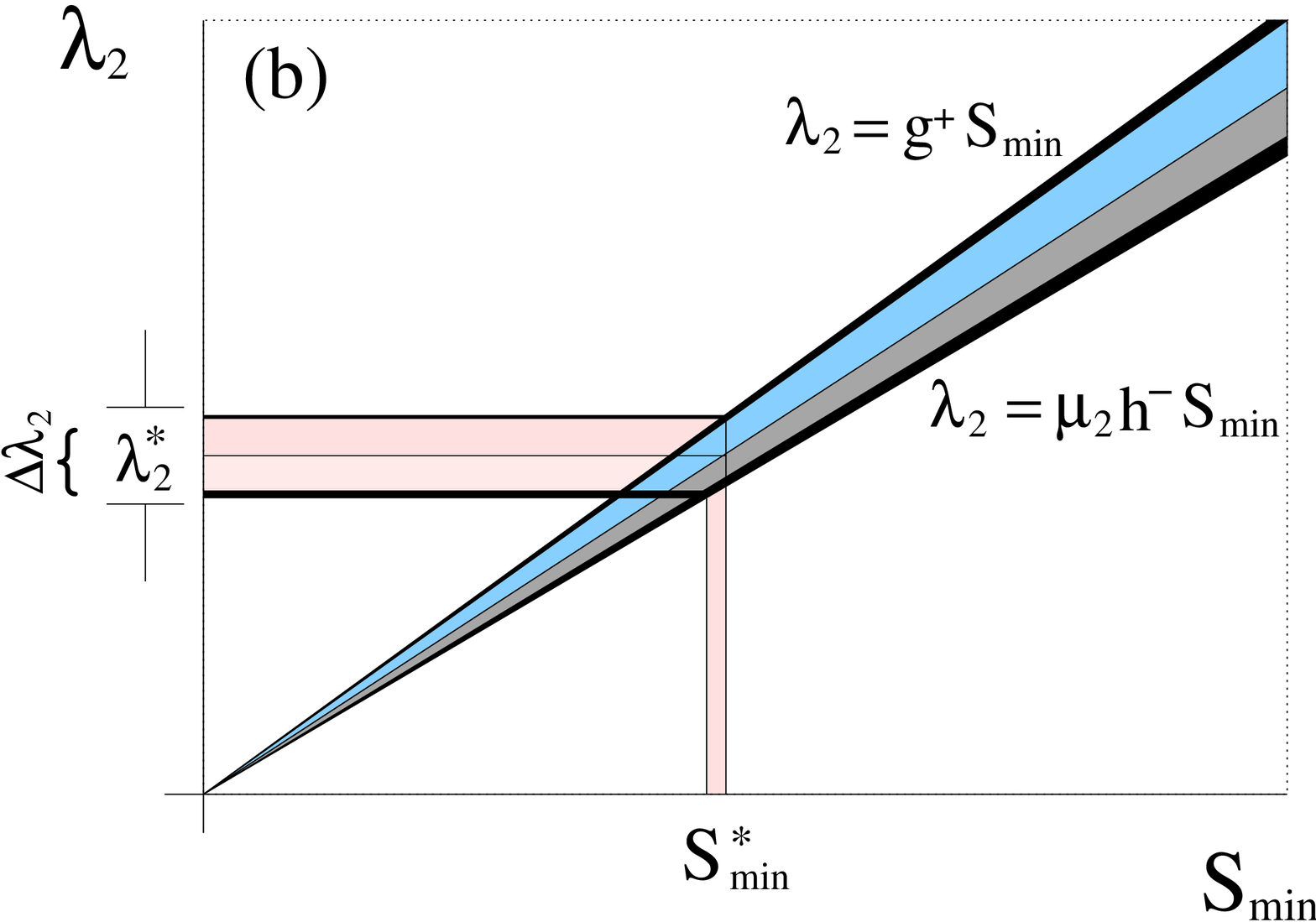}
\caption{(a) Illustration of the design of networks with extreme Laplacian eigenvalues 
(a) $\lambda_N^*\in [S^*_{\max}, S^*_{\max} \times \mu_N]$ 
and 
(b) $\lambda_2^*\in [S^*_{\min} \times \mu_2 h, S^*_{\min} \times g]$  
for a given substrate network with normalized eigenvalues $\mu_N$ and $\mu_2$.
In (a), for given $S_{\max}$, the blue area (bottom) represents the uncertainty in
$\lambda_N$
due to the factor $\mu_N$ and the triangular line (top) accounts for any inaccuracy in the
determination of $\mu_N$. In (b),  for given $S_{\min}$, the black, blue, gray and black areas 
(top to bottom) represent the uncertainty in $\lambda_2$ due to the factors $g$, $\mu_2$, $h$,
and any inaccuracy in the determination of $\mu_2$, respectively. Here,
$g^+$ and $h^-$ are  used to indicate the maximum and 
minimum of $g$ and $h$, respectively, in the given interval of $S_{\min}.$
}
\label{fig1}
\end{center}
\end{figure}

Given an arbitrary {\it substrate} network of $N$ nodes and known eigenvalues $\mu_N$ and
$\mu_2$, the bounds in Eqs.~(\ref{eq41}) and (\ref{eq42}) can be used to generate networks
of
 eigenvalues $\lambda_N^*=\lambda_N\pm \Delta\lambda_N$ and $\lambda_2^*=\lambda_2 \pm \Delta\lambda_2$, 
where the uncertainties $\Delta\lambda_N$ and $\Delta\lambda_2$ depend on $|\mu_N-1|$
and $|g-\mu_2 h|$, respectively. Here, $\lambda_N$ and $\lambda_2$ denote
the desired values and $\lambda_N^*$ and $\lambda_2^*$ denote the resulting eigenvalues, which have
some uncertainty.
This procedure is illustrated in Fig.~\ref{fig1}
and can be used to systematically design robust networks with tunable extreme eigenvalues.

The rationale here is that the substrate network is chosen to incorporate topological
constraints relevant to the problem, such as the nonexistence of links between certain
nodes or a limit in the number of links, and that the extreme eigenvalues of
the normalized Laplacian $\hat{G}$ of this network are calculated beforehand. Then, by
adjusting the minimum and maximum input strengths $S_{\min}$ and $S_{\max}$, one can
define new networks with the same topology but with the desired extreme eigenvalues
for the Laplacian matrix $G$. 

More specifically, if $\mu_N$ is known, one can adjust the largest input strength using
Eq.\ (\ref{eq41}) to
obtain a new network with $\lambda_N^*$ in the interval $[\lambda_N, \lambda_N \times \mu_N]$
by setting  $S_{\max}=S^*_{\max}\equiv\lambda_N$ [see Fig.~\ref{fig1}(a)]. 
Likewise, if $\mu_2$ is known, one can use Eq.\ (\ref{eq42}) to adjust the minimum input
strength and
generate a new network with $\lambda_2^*$ within a desirable interval 
$[\lambda_2 \times \mu_2 h, \lambda_2 \times g]$ by taking $S_{\min}=S^*_{\min}\equiv \lambda_2$ 
[see Fig.~\ref{fig1}(b)]. Naturally, the usefulness of this construction will depend on how close
to $1$ are the eigenvalue $\mu_N$ and $\mu_2$, and how close to $1$ are kept $h$ and $g$ as the
weights are changed.  The former condition
can be justified for
most networks in the usual ensembles of
densely connected random networks
and also 
in ensembles of sparse networks with large mean degree $k$
\cite{chung:book}. 
Note that this approach can be effective even when $\mu_N$ and $\mu_2$ are only approximately known,
as represented by the upper and lower black diagonal lines in Fig.~\ref{fig1}(a) and
(b), respectively. The last observation is relevant precisely when $\mu_N$ and $\mu_2$ are
estimated from an ensemble distribution or through any other probabilistic procedure.

Importantly, because $\lambda_2$ is mainly controlled by $S_{\min}$ and $\lambda_N$ by $S_{\max}$, 
both eigenvalues can be adjusted simultaneously. In the synchronization problem, this can be used to
define networks with predetermined synchronizability $R$ and predetermined upper bound for the coupling
cost $C$. Moreover, this construction is not unique, that is, there are multiple choices of the substrate
network and of the assignment of weights $\{S_i\}_{i=1}^{N}$ versus degrees  $\{k_i\}_{i=1}^{N}$ that
will lead to the same pair of predefined eigenvalues $\lambda_2$ and $\lambda_N$. This freedom can
be explored to increase robustness against structural perturbations and to control the uncertainty
by keeping $h$ large and $g$ small.

\begin{figure}[t]
\begin{center}
\includegraphics[width=0.48\textwidth]{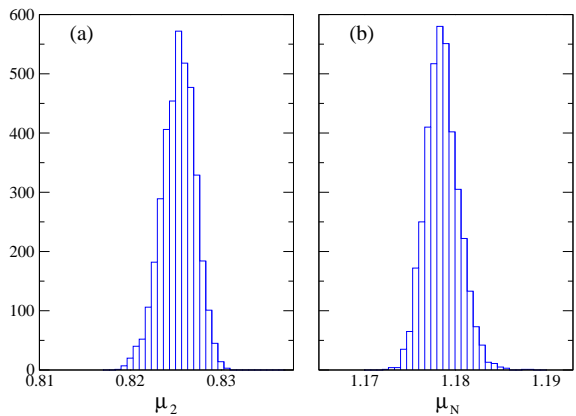}
\caption{Distribution of the eigenvalues (a) $\mu_2$ and (b) $\mu_N$
for Erd\H{o}s-R\'enyi networks. The histrograms correspond to $3800$
realizations of the networks for $N=500$ and $p=0.2$.
}
\label{fig2}
\end{center}
\end{figure}

\subsection{Numerical Example}

Consider unweighted Erd\H{o}s-R\'enyi networks, generated by adding 
with probability $p$ a link between each pair of $N$ given nodes \cite{er}. 
As shown in the histograms of Fig.\ \ref{fig2}, the  eigenvalues 
$\mu_2$ and $\mu_N$ are narrowly distributed close to $1$ even for
relatively small and sparse networks. Such networks can thus be used as
substrate networks to generate, with good accuracy, new networks of predefined
eigenvalues $\lambda_2$ and $\lambda_N$ by reassigning the input strengths
$S_{\min}$ and $S_{\max}$, respectively. 

While a single realization of the substrate network and a deterministic assignment
of input strengths $S_{\min}$ and $S_{\max}$ would suffice to generate the desired
networks, the robustness of the proposed procedure becomes more visible if one considers 
various independent random constructions. For this purpose, I consider
random realizations of the substrate network and assume that, for each such realization,
the input strength of each node is assigned with equal probability to be either $S_{\min}$
or $S_{\max}$.

\begin{figure}[t]
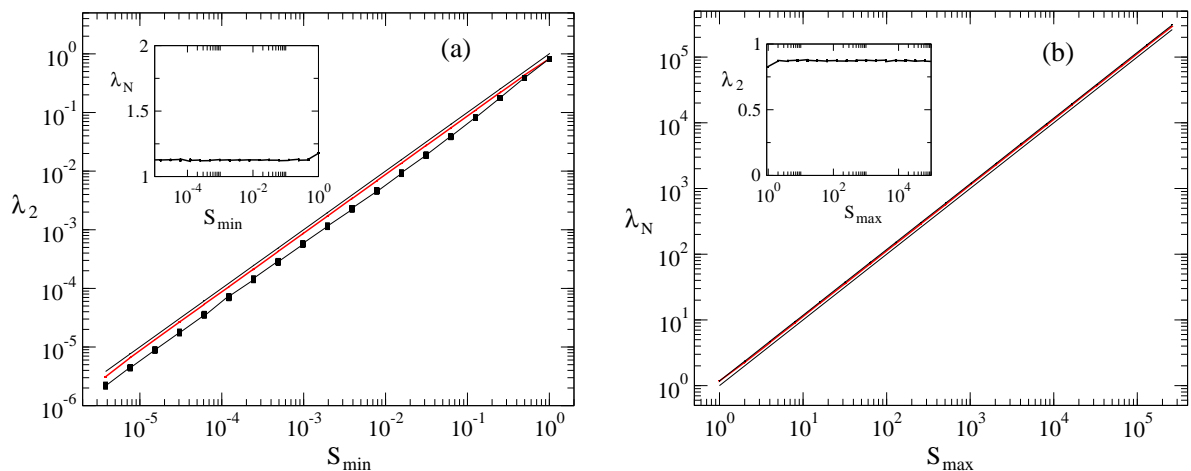

\begin{center}
\includegraphics[width=0.48\textwidth]{sfig1.eps}\hspace{0.5cm}
\includegraphics[width=0.48\textwidth]{sfig2.eps}
\caption{Design of networks with tunable eigenvalues
(a) $\lambda_2$ and (b) $\lambda_N$:
the black lines indicate the upper and lower bounds given by Eqs.\ (\ref{eq41})-(\ref{eq42})
and the red lines indicate to the numerically determined eigenvalues as functions of $S_{\min}$
(for $S_{\max}=1$) and of $S_{\max}$ (for $S_{\min}=1$), respectively. 
Each choice of $S_{\min}$ in (a) and of $S_{\max}$ in (b) corresponds to $100$ realizations of 
the substrate networks for the same parameters used in Fig.\ \ref{fig2}. Insets: distributions
of (a) $\lambda_N$ and (b) $\lambda_2$ for the networks used in the main panels 
(a) and (b), respectively.
}
\label{fig3}
\end{center}
\end{figure}

Figure \ref{fig3} shows the numerically computed eigenvalues $\lambda_2$ and $\lambda_N$,
and the respective bounds, as functions of $S_{\min}$ and $S_{\max}$. This figure
is a scattered plot with $100$ independent realizations of the substrate
networks (and assignments of input strengths) for each choice of $S_{\min}$ and 
$S_{\max}$. As shown in the figure, except for the lower bound of
$\lambda_2$, which exhibits observable dependence on the specific network realization, the 
distributions of the eigenvalues and  bounds are narrower than the width of the lines
in the figure. In addition, the numerically computed values of $\lambda_2$ and $\lambda_N$ are tightly
bounded by the lower and upper limits in Eqs.\ (\ref{eq41})-(\ref{eq42}). The difference between the 
bounds of $\lambda_N$ in Fig.\ \ref{fig3}(b) is thinner than the width
of line. Moreover, as $S_{\min}$ ($S_{\max}$) is varied for fixed $S_{\max}$ ($S_{\min}$) 
in Fig.\ \ref{fig3}(a) (Fig.\ \ref{fig3}(b)) , 
the value of $\lambda_N$ 
($\lambda_2$) remains nearly constant, as shown in the insets. 
Thus, by varying both $S_{\min}$ and $S_{\min}$, one can design networks where
both $\lambda_2$ and $\lambda_N$ are predetermined. 

\begin{figure}[h]
\begin{center}
\includegraphics[width=0.48\textwidth]{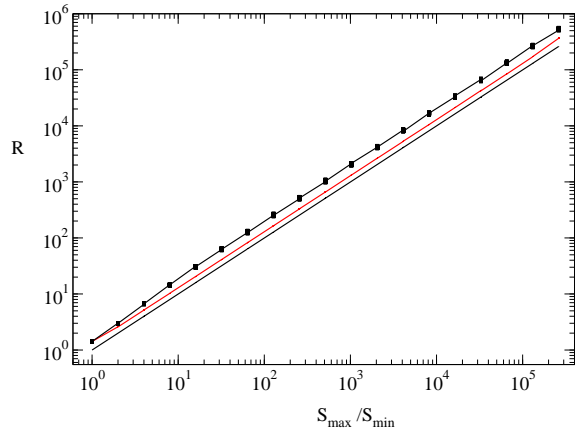}
\caption{Same as in Fig.\ \ref{fig3} for the ratio $R=\lambda_N/\lambda_2$ as a function of
the ratio $S_{\max}/S_{\min}$.}
\label{fig4}
\end{center}
\end{figure}

Figure \ref{fig4}, shows the result of such a construction for the ratio of eigenvalues
$R$. Note that if all the input strengths $S_i$ are re-scaled by a common factor $\alpha$, 
the terms $\nu_N$, $\lambda_N$, and $\nu_N\mu_N$ in Eq.\ (\ref{eq41}) as well
as the terms $\nu_1\mu_2 h$, $\lambda_2$, and  $\nu_1 g$ in Eq.\ (\ref{eq42}) will change
by the same factor $\alpha$. Therefore, the ratio $R$ and corresponding bounds do not
change if, in our simulations,  both $S_{\min}$ and $S_{\max}$ are re-scaled by a common
factor.

\medskip

\noindent
{\it Remark:} If no constraints are imposed to the topology of the network other than the number $N$
of nodes, then one could easily construct networks having exactly any given set of eigenvalues
$0=\lambda_1<\lambda_2\le \dots \le \lambda_N$ and any given set of orthonormal eigenvectors 
$u_1, \cdots, u_N$, where $u_1^T=(1/\sqrt{N},\dots, 1/\sqrt{N})$. The network satisfying
this conditions is defined by the symmetric Laplacian $G= UdU^T$, where $d$ is the diagonal matrix
of eigenvalues $\{\lambda_i\}_{i=1}^N$ and $U$ is the orthogonal matrix of eigenvectors $\{u_i\}_{i=1}^N.$
Note that matrix $G$ is indeed a well-defined Laplacian satisfying the zero row sum condition.

\section{Impact of the Network Structure on Synchronization}
\label{sec5}

Equations (\ref{eq41}) and (\ref{eq42}) can be used to address the influence of 
the network structure on the dynamics. In particular, they imply that
\begin{equation}
\frac{S_{\max}}{S_{\min}}\frac{1}{g}\le R\le\frac{S_{\max}}{S_{\min}}\frac{\mu_N}{\mu_2}\frac{1}{h}.
\label{e11}
\end{equation}
Therefore, under rather general conditions, the synchronizability of the network is strongly limited by 
$S_{\max}/S_{\min}$ and $\mu_N/\mu_2$. The first ratio depends on the distribution of weights
while the second also depends on the topology of the network. 
The bounds in Eq.~(\ref{e11}) are valid for any network satisfying condition (\ref{e3}), but are
tighter for classes of networks with $\mu_N/\mu_2$, $g$ and $h$ closer to $1$. In this section 
I focus on large random networks, which forms one such class of networks.

\subsection{Synchronizability of Random Networks}

For concreteness, consider random networks for which the normalized matrix 
$\hat{A}$ is unweighted. That is, random networks which are either unweighted 
or whose weights are factored out completely in Eq.~(\ref{e3}). 
For these networks, one can invoke the known result from graph spectral theory
\cite{chung:book} that the expected values of the extreme eigenvalues of $\hat{G}$
approach $1$ as
$\langle\mu_N\rangle = 1 + O(1/\sqrt{k})$ and 
$\langle\mu_2\rangle = 1 - O(1/\sqrt{k})$ for large mean degree $k$.  
This behavior has been shown to remain valid for networks with quite general expected degree
sequence \cite{CLV:2003} and to be consistent with
numerical simulations on various models of growing and scale-free networks \cite{MCK1,MCK2},
even when the networks are relatively small and only approximately random insofar as $k_{\min}\gg 1$.
In addition, the distribution of the eigenvalues across the ensemble of random networks becomes
increasingly peaked around the expected values as the size of the networks increases \cite{CLV:2003,DH07}.
Furthermore, for most realistic networks, $h$ is bounded away from zero
and $g$ approaches $1$ for large $N$ [it can be replaced by $1$ if the conditions in {\it remark 1} (appendix)
apply]. For unweighted networks, in particular,  
$h= (k \bar{k}^{-1})^{-1/2}$ and $g=N/(N-1)$, where $\bar{k}^{-1}$ is the average of $1/k_i$ in the network.
Therefore, for a wide class of complex networks, the eigenvalues $\lambda_N$ and $\lambda_2$ are
mainly determined by $S_{\max}$, through $S_{\max}\le \lambda_N\le S_{\max}\mu_N$, and by 
$S_{\min}$, through $S_{\min} \mu_2 h\le \lambda_2\le S_{\min}g$, respectively. 

In the case of unweighted (and undirected) networks, the input strengths are determined by the
degrees of the nodes and $S_{\max}/S_{\min}=k_{\max}/k_{\min}$. Thus, the bounds in
Eq.~(\ref{e11}) can be used to assess the effect of the degree distribution.
As a specific example, consider random scale-free networks 
\cite{barab,semi_random} 
with degree distribution
$P(\kappa)= c \kappa^{-\gamma}$ for $\kappa\ge k_{\min}$ and $\gamma>2$, where
$1/c =\sum_{\kappa=k_{\min}}^{N-1}\kappa^{-\gamma}\approx k_{\min}^{-\gamma+1}/(\gamma-1)$
is a normalization factor. From the condition $N\int_{k_{\max}}^{\infty}P(\kappa)d\kappa=1$ \cite{no_cutoff},
one has $k_{\max}/k_{\min}\approx N^{1/(\gamma-1)}$, which leads to
\begin{equation}
R \sim N^{1/(\gamma-1)}
\label{e13}
\end{equation}
for large $N$ and $k_{\min}$ \cite{diff_scalings}. This simple scaling for the expected value
of $R$ explains the counter-intuitive
results about the suppression of synchronizability in networks with heterogeneous
distribution of degrees reported in Ref.~\cite{NMLH:2003}. Random scale-free
networks were found to become less synchronizable as the scaling exponent
$\gamma$ is reduced, despite the concomitant reduction of the average distance
between nodes \cite{cohen} that could facilitate the communication between the synchronizing
units \cite{NMLH:2003}. Equation~(\ref{e11}) shows that
this effect of the degree distribution is a direct consequence of the increase in
the heterogeneity of the input strengths, characterized by $S_{\max}/S_{\min}=k_{\max}/k_{\min}$.
Equation~(\ref{e13}) predicts
this effect as a function of both the scaling exponent $\gamma$ and the size $N$
of the network. In particular, this equation shows that scale-free networks become more difficult
to synchronize as $N$ increases and this is again because $S_{\max}/S_{\min}
\approx N^{1/(\gamma-1)}$ increases. On the other hand, synchronizability increases
as $\gamma$ is increased and becomes independent of the system size for $\gamma=\infty$,
indicating that networks with the same degree for all the nodes are the most
synchronizable random unweighted networks (see also Ref.\ \cite{expanders}).

In the more general case of weighted networks, the input strengths are not
necessarily related to the degrees of the nodes. An important implication of
Eq.~(\ref{e11}) is that, given a heterogeneous distribution of input strengths $S_i$ in Eq.~(\ref{e3}),
the synchronizability of the network is to some extent independent of the way the input
strengths are assigned to the nodes of the network, rendering essentially the
same result whether this distribution is correlated or not with the degree
distribution. In both cases, synchronizability is mainly determined by the
heterogeneity of the input strengths $S_{\max}/S_{\min}$ and the mean degree $k$. 
In particular, synchronizability tends to be
enhanced (suppressed) when the mean degree $k$ is increased (reduced) and when
the ratio $S_{\max}/S_{\min}$ is reduced (increased). This raises the interesting
possibility of controlling the synchronizability of the network by adjusting these
two parameters, which was partially explored in Sec.~\ref{sec4}.

\subsection{Structural Control of the Dynamics}

As a specific example of control, consider a given random network with arbitrary
input strengths $\{S_i\}_{i=1}^N$, where the topology of the network is kept fixed
and the input strengths are redefined as
\begin{equation}
S_i'(\theta) = (S_i)^{\theta},
\label{e14}
\end{equation}
with $\theta$ regarded as a tunable (control) parameter. For large $k$,
synchronizability is now mainly determined by
$\max_{i,j}\{S_i'(\theta)/S_j'(\theta)\}= (S_{\max}/S_{\min})^{\theta}$.
Within this approximation,  synchronizability is expected to reach
its maximum around $\theta=0$, quite independently of the 
initial distribution of input strengths $S_i$ and the details of the degree
distribution. This generalizes a result first announced in Ref.~\cite{MCK1},
namely that networks with good synchronization properties tend to be at least
approximately uniform with respect to the strength of the input signal
received by each node (but see remark below).

These optimal networks have interesting properties. For $\theta=0$,
all the nodes of the network have exactly the same input strength. 
Thus, if nodes $i$ and $j$ are connected, the strength of the connection from
$j$ to  $i$ scales as $1/k_i$, while the strength of the connection from
$i$ to  $j$ scales as $1/k_j$.  This indicates that, unless all the nodes
have exactly the same degree, the networks that optimize synchronizability
for that degree distribution are necessarily  weighted {\it and} directed.
Moreover, if $k_j>k_i$, the
strength $\propto 1/k_i$ of connection from node $j$ to node $i$ is larger than the
strength $\propto 1/k_j$ of the connection from node $i$ to node $j$. Therefore,
in the most synchronizable networks, the dynamical units are asymmetrically
coupled and the stronger direction of the connections is from the nodes with
higher degrees to the nodes with lower degrees. The asymmetry and the predominance
of connections from higher to lower degree nodes is a consequence of the condition
that nodes with different degrees have the same input strength, a condition that 
introduces correlations between the weights of individual connections and the topology
of the network and that has been observed to have similar consequences in 
other coupling models \cite{MCK3,b:2006}.
These results combined with the interesting recent work of Giuraniuc {\it et al.}
\cite{Giuraniuc} on critical behavior suggest that, in realistic systems, the
properties of individual connections are at least partially shaped by the
topology of the network.

\medskip

\noindent{\it Remark:}
The above analysis shows that for the networks satisfying the condition in Eq.~(\ref{e3}), 
$R$ is more tightly bounded close to the optimal value $R=1$ when the distribution of input strengths
$S_i$ is more homogeneous. Indeed, the bounds in Eq.~(\ref{e11}) leave little room for the improvement
of synchronizability by changing the weights of individual links or the way the nodes are
connected if  $S_{\max}/S_{\min}$ is not reduced. 
For classes of more general directed networks, however,  one can have
highly synchronizable networks with a heterogeneous distribution of $S_i$. To see this,
consider the set of most synchronizable networks among all possible networks, which is
precisely the set of networks with $R=1$ and eigenvalues
$\lambda_2 = \cdots \lambda_N\equiv \lambda>0$. As shown in
Refs.~\cite{TM2,TM1}, if the Laplacian matrix $G$ is diagonalizable, then the networks
with $R=1$ are those where each node either has output connections with the same strength
to all the other nodes (and at least one node does so) or has no output connections at all.
From this and the zero row sum property of the Laplacian matrix, it follows that $S_i=\sum_{j\neq i} a_j
=\lambda -a_i$, where $a_i\ge 0$ is the strength of each output connection from node $i$. 
Accordingly, the input strength $S_i$ is upper bounded by $\lambda$, but not necessarily
the same for all the nodes. In particular, since the strengths of the output connections can
have any values $a_i\ge 0$ (as long as at least one is non-zero), in this case there is no lower
limit for $S_i$ and the ratio $S_{\max}/S_{\min}$ can be arbitrarily large 
despite the fact that $R=1$. 
Therefore, even when the spectra is real, strictly directed networks can be fundamentally
different from the directed networks considered here \cite{outros}. 


\section{Concluding Remarks}

I have presented rigorous results showing that the extreme eigenvalues of the Laplacian 
matrix of many complex networks are bounded by the node degrees and input strengths, where
the latter can be interpreted as the {\it weighted} in-degrees in the networks. These
results can be used to predict and control the coupling cost and a number of implications of
the network structure on the dynamical properties of the system, 
such as its tendency to sustain synchronized behavior. I have shown here that these results
can also be used to design networks with predefined dynamical properties.  

While I have focused on complete synchronization of identical units, the leading role of 
$S_{\max}/S_{\min}$ and $k$ revealed in this work also provides insights into other
forms of synchronization. In particular, it seems to help explain: the suppressive effects of
heterogeneity in the synchronization of pulse-coupled \cite{DTDWG:2004} and non-identical
oscillators \cite{MCK3}; the dominant effect of the mean degree in the synchronization of
time-delay systems with normalized input signal \cite{MM}; and the dominant effect of the degree
in the synchronization of homogeneous networks of bursting neurons \cite{BLH}. 
The scale-free
model of neuronal networks considered in Ref. \cite{GL2005}, which was shown to generate large
synchronous firing peaks, is also consistent with (an extrapolation of) the results above. Indeed, the
networks in that model are scale free only with respect to the out-degree distribution and are
homogeneous with respect to the in-degree distribution. 
Therefore, the results presented here may serve as a reference in the study of more general systems,
including those with heterogeneous dynamical units \cite{costa, nid1, nid2, nid3, nid4}.
In general, the impact of the network structure will change both with the specific synchronization model
and with the specific question under consideration, and an important open problem is to understand {\it how}
it changes.

Finally, since the Laplacian eigenvalues also govern a variety of other processes \cite{spec1,spec2},
including the relaxation time in diffusion dynamics \cite{MCK2}, community formation \cite{arenas:2006}, 
consensus phenomena \cite{kevin},
and first-passage time in random walk processes \cite{DHM}, 
the results reported here are also expected to meet other applications in the broad area
of dynamics on complex networks, particularly in connection with network design in communication 
and transport problems.

\subsection*{Acknowledgments}

The author thanks Dong-Hee Kim for valuable discussions and for reviewing the manuscript.


\appendix
\section*{Appendix. Proof of the Theorem}
\setcounter{section}{1}

In what follows I use the notation that, if $X$ is a $N \times N$ matrix with eigenvalues
$\alpha_1\le \alpha_2\le \cdots \le \alpha_N$, then $v_i^X$ denotes a normalized
eigenvector of eigenvalue $\alpha_i$.  The proof of the theorem is divided in 6 steps.\\

\noindent{\it Step 1}: The eigenvalues of matrices  $\hat{G}$ and $G$ satisfy
\begin{eqnarray}
\mbox{eig}(\hat{G}) &=& \mbox{eig}(H), 
\label{eq61}\\
\mbox{eig}(G) &=& \mbox{eig}(Q),
\label{eq62}
\end{eqnarray}
where $H=D^{-1/2}LD^{-1/2}$ and $Q=S^{1/2}HS^{1/2}$. 
Equations ~(\ref{eq61}) and  (\ref{eq62}) 
follow from the identities det$(\hat{G}-\alpha I)=$ det$(H-\alpha I)$ and 
det$(G-\alpha I)=$ det$(Q-\alpha I)$, respectively,
where $\alpha$ is an arbitrary number
and $I$ is the $N\times N$ identity matrix. 
Because matrices $H$ and $Q$ are symmetric, their eigenvalues are real,
as assumed in Eqs.~(\ref{eq31}) and (\ref{eq32}), and the corresponding
eigenvectors can be chosen to form orthonormal bases \cite{num}.  \\

\noindent{\it Step 2}: The diagonalizability of matrices $G$ and $\hat{G}$,
a condition invoked in the rest of this appendix, can be demonstrated as
follows. Matrix $Q$ is symmetric and hence has a set of orthonormal eigenvectors
$\{v_i^Q\}_{i=1}^{N}$. Then, from the identity $G=S^{1/2}D^{-1/2}QS^{-1/2}D^{1/2}$,
and the fact that $S^{1/2}D^{-1/2}$ is nonsingular, it follows that
$\{S^{1/2}D^{-1/2}v_i^Q\}_{i=1}^{N}$ forms a set of $N$ linearly independent
eigenvectors of $G$. This implies that $G$ is diagonalizable. From the
special case $S=I$, it follows that the same holds true for $\hat{G}$.\\

\noindent{\it Step 3}: The upper bound of $\lambda_N$ in Eq.~(\ref{eq41})
follows immediately from 
\begin{equation}
\lambda_N=\max_{\|v\|=1}\|S\hat{G} v\| \le \max_{\|v\|=1}\|Sv\| \max_{\|v\|=1}\|\hat{G} v\|
\label{eq7}
\end{equation}
where $\| . \|$ is the usual Euclidean norm.\\

\noindent{\it Step 4}: The lower bound of $\lambda_N$ in Eq.~(\ref{eq41})
is derived from
\begin{equation}
\lambda_N=\max_{\|v\|=1}\|Gv\|\ge\|G\bar{v}\|,
\label{eeq8}
\end{equation}
where $\bar{v}$ is a unit vector chosen such that $\bar{v}_i=\delta_{ij(N)}$
and $j(N)$ is the index of a node with the largest input strength.
Equation (\ref{eeq8}) leads to
\begin{equation}
\lambda_N^2\ge S_{j(N)}^2+\sum_{i} \hat{A}^2_{ij(N)}(S_i/k_i)^2,
\label{eqn1}
\end{equation}
and this leads to the lower bound in Eq.~(\ref{eq41})
with a strict inequality for finite size networks.
In the particular case of unweighted
networks, Eq.~(\ref{eqn1}) implies $\lambda_N\ge k_{\max} \sqrt{1+1/k_{\max}}$
(see also Ref.\  \cite{grone_merris}).\\

\noindent{\it Step 5}:
Now I turn to the upper bound of $\lambda_2$ in Eq.~(\ref{eq42}).
From the identity eig$(G)=$ eig$(Q)$, one has
\begin{equation}
\lambda_2=\min_{\|v\|\ne 0 \;|\; v\bot v_1^Q}\frac{\langle v,Qv\rangle}{\langle v, v\rangle},
\label{step5}
\end{equation}
where $\langle . , . \rangle$ denotes the usual scalar product.
This equation can be rewritten as
\begin{equation}
\lambda_2 =
\sum_j (k_j/S_j) \min_{\|v\|\ne 0 \;|\; v\nparallel v_1^Q}\frac{\sum_{i,j}\hat{A}_{ij}(\sqrt{S_i/k_i}v_i - \sqrt{S_j/k_j}v_j)^2}
{\sum_{i,j}(\sqrt{k_j/S_j}v_i - \sqrt{k_i/S_i}v_j)^2},
\label{eeq9}
\end{equation}
where I have used that $v_1^Q\propto (\sqrt{k_1/S_1}, \cdots ,\sqrt{k_N/S_N})$
to obtain the identities
\begin{eqnarray}
\sum_{i,j}( \sqrt{k_j/S_j}v_i - \sqrt{k_i/S_i}v_j)^2 =2\sum_j( k_j/S_j ) \langle v^{\bot},v^{\bot} \rangle,\nonumber\\
\sum_{i,j}\hat{A}_{ij}(\sqrt{S_i/k_i}v_i - \sqrt{S_j/k_j}v_j )^2=2\langle v^{\bot},Qv^{\bot} \rangle,\nonumber
\end{eqnarray}
where $v^{\bot}$ is the component of $v$ orthogonal to $v_1^Q$.
The minimum in Eq.~(\ref{eeq9}) can be upper-bounded by taking $v_i=\delta_{ij(1)}$,
where $j(1)$ is the index of a node with the smallest input strength, and this leads to 
the upper bound in Eq.~(\ref{eq42}).\\

\noindent{\it Remark 1:} A different bound,
$\lambda_2\le (S^2_{j'}k_{j''}+S^2_{j''}k_{j'}+2\hat{A}_{j'j''}S_{j'}S_{j''})/(S_{j'}k_{j''}+S_{j''}k_{j'})$,
is obtained for any $j''\neq j'$ by using 
$\bar{v}_i=\sqrt{S_{j'}/k_{j'}}\delta_{ij'} - \sqrt{S_{j''}/k_{j''}}\delta_{ij''}$
to upper-bound $\lambda_2$ in Eq.~(\ref{step5}). This leads to $\lambda_2\le \nu_1$
if there are two nodes with minimum input strength $S_{\min}$ 
that are not connected to each other.\\

\noindent{\it Remark 2:} Alternatively, one can show that eig$(G)=$ eig$(H^{1/2}SH^{1/2})$ and use
this to
upper-bound $\lambda_2$ with $\| H^{1/2}SH^{1/2}v\|/\|v\|$ for $v\bot v_1^H$ in the span
of $\{ v^S_1, v^S_2\}$. If there are two or more nodes with minimum input strength $S_{\min}$,
then it follows from this bound that $\lambda_2\le \nu_1\mu_N$.\\

\noindent{\it Step 6}: The lower bound of $\lambda_2$ in Eq.~(\ref{eq42}) is derived as follows.
From the identity eig$(G)=$ eig$(Q)$, one has
\begin{eqnarray}
\lambda_2 = \min_{\|v\|=1 \;|\; v\bot v_1^Q}\| Qv\| 
          = \min_{\|v\|=1 \;|\; v\bot S^{-1/2}v_1^H}\| S^{1/2}HS^{1/2}v\|. \nonumber
\label{eq17}
\end{eqnarray}
The identity
\begin{equation}
\| S^{1/2}HS^{1/2}v\|=  \left\| S^{1/2} \frac{HS^{1/2}v}{\|HS^{1/2}v\|}\right\|
                        \left\| H \frac{S^{1/2}v}{\|S^{1/2}v\|}\right\| \left\| S^{1/2}v\right\|\nonumber
\end{equation}
and the observation that the minimum of the product is lower-bounded by the product of the minimums
lead to
\begin{equation}
\lambda_2  \ge  \nu_1 \min_{\|v\|\ne 0 \;|\; v\bot v_1^H} \left\| H \frac{Sv}{\|Sv\|}\right\|
           \ge   \nu_1\mu_2 \| v^{H\bot}\|,
\label{eq18}
\end{equation}
where 
\begin{equation}
\|v^{H\bot}\|^2 = 1 - \max_{\|v\|\ne 0 \;|\; v\bot v_1^H} \frac{|\langle Sv,v_1^H\rangle|^2}{\|Sv\|^2}.
\label{eq22}
\end{equation}
In the r.h.s.~of Eq.~(\ref{eq22}) one has the maximum of function $|\langle Sv,v_1^H\rangle|^2$ under the
constraints $\langle v, v_1^H \rangle=0$ and $\|Sv\|=1$, which can be determined using the Lagrange Multipliers
Method with two multipliers. The resulting set of equations is
\begin{eqnarray}
\sum_i\sqrt{k_i}v_i &=& 0,\\
\sum_iS_i^2v_i^2    &=& 1,\\
\frac{S_i\sqrt{k_i}}{\sqrt{kN}} &=& m_1\frac{\sqrt{k_i}}{\sqrt{kN}}+m_2 S_i^2 v_i,
\label{eq23}
\end{eqnarray}
where $m_1$ and $m_2$ are the Lagrange Multipliers. This system of equations
can be solved for $m_2=\max |\langle Sv,v_1^H\rangle|$ under the corresponding constraints
by taking $\sum_i$ of Eq.~(\ref{eq23}) multiplied by $v_i$, $\sqrt{k_i}/S_i$, and $\sqrt{k_i}/S_i^2$,
respectively. The result is
\begin{equation}
m_2^2= 1-\frac{(\sum_i k_i/S_i)^2}{(\sum_i k_i)(\sum_i k_i/S_i^2)}.
\label{eq24}
\end{equation}
The lower bound in Eq.~(\ref{eq42}) follows from Eqs.~(\ref{eq18}), (\ref{eq22}), and (\ref{eq24}),
and this concludes the proof of the theorem.
$\Box$


\begin{thebibliography}{99}
\bibitem{num} An important byproduct of identities (\ref{eq61})  and (\ref{eq62}) is that, numerically,
the computation of the eigenvalues $\mu_i$ and $\lambda_i$ is significantly less time demanding when
evaluated from the symmetric matrices $H$ and $Q$, respectively.
\end{thebibliography}
\end{document}